# Distributed Fiber Monitoring on a Coherent DSP ASIC: Implementation of Longitudinal Power Monitoring

Takeo Sasai, *Senior Member, IEEE*, Yukinobu Nakajima, Minami Takahashi, Runa Kaneko, Masanori Nakamura, *Member, IEEE*, Asuka Matsushita, Fukutaro Hamaoka, *Member, IEEE*, and Etsushi Yamazaki, *Member, IEEE*

(*Post-Deadline Paper*)

***Abstract*— Coherent DSP ASICs and transceivers to date report a rich set of optical link metrics, yet these remain cumulative over the link and provide no spatial localization. This paper presents the implementation design of fiber-longitudinal power monitoring (LPM)—which recovers spatially resolved power profiles from received traffic—on a coherent DSP ASIC chip. We first identify the challenges specific to DSP implementation, and describe the enabling algorithm and architecture. We then present the key design considerations for retaining the estimation fidelity of the on-chip LPM during in-service operation. On this basis, we implement LPM on a coherent DSP—a 3-nm 800G low-power DSP ASIC housed in an OSFP pluggable module—which constitutes, to our knowledge, the first coherent DSP ASIC and transceiver to perform distributed fiber monitoring and data communication simultaneously. The on-chip LPM is experimentally validated over links of up to 1005 km using 400ZR+ and 800ZR+ traffic, with no measurable impact on Q factor or module power consumption. Multi-vendor interoperability is also demonstrated, where the on-chip LPM operates on signals transmitted by third-party transceivers. These results extend coherent DSPs from cumulative to spatially resolved monitoring, allowing pluggable transceivers—deployable from DCI to long-haul—to act not only as cost-efficient transmission elements but also as distributed monitoring points across diverse optical network infrastructure.**

## I. Introduction

Over the past fifteen years, the digital signal processing (DSP) application-specific integrated circuit (ASIC) has drastically advanced the data rate and reach of coherent transmission while reducing power consumption, cost, and footprint [1]. This relentless reduction in power and footprint, driven by both CMOS process scaling and DSP algorithmic advances, has brought coherent optics—once confined to bulky dedicated line cards—into compact pluggable modules, progressing from CFP2 to the more power-efficient QSFP-DD and OSFP form factors. As this efficiency improved, power-hungry long-reach coherent transmission has itself moved into the compact pluggable form factor [2]: though initially targeted for data center interconnect (DCI) as in 400ZR [3], coherent pluggables have since been expanding to multi-span metro and long-haul, as exemplified by ZR+ [4][5]. Consequently, pluggable transceivers are increasingly serving as a common transmission platform covering the entire range of optical networks, which is one of the motivations for integrating longitudinal power monitoring (LPM) into the low-power coherent DSP ASIC for widespread adoption.

These pervasively deployed real-time coherent DSP ASICs already provide more than data transport, though in a limited form. State-of-the-art DSPs have link monitoring capabilities, reporting various physical metrics extracted from their signal compensation blocks: chromatic dispersion (CD), differential group delay (DGD), state of polarization (SOP), polarization-dependent loss (PDL), and signal quality metrics [6][7], all of which are accessible from coherent pluggables as defined in the Coherent Common Management Interface Specification (C-CMIS) [8]. These metrics represent cumulative values with no spatial or longitudinal resolution along the transmission fiber. Spatially resolved characterization is instead obtained with separate hardware devices such as optical time-domain reflectometry (OTDR), which are well established yet typically require dedicated instrumentation and associated operational overhead.

An alternative means of obtaining spatially resolved information is LPM, also referred to as optical link tomography, which derives distributed power profiles along a fiber link by processing communication signals received at a coherent receiver [9][10]. By leveraging the interaction between Kerr nonlinearity and CD in fibers, LPM achieves forward, distributed fiber monitoring: unlike most fiber-sensing techniques, it requires neither dedicated interrogation hardware

T. Sasai, Y. Nakajima, M. Takahashi, R. Kaneko, M. Nakamura, A. Matsushita, F. Hamaoka, and E. Yamazaki, are with the Network Innovation Laboratories, NTT, Inc., Kanagawa 239-0847, Japan (e-mail: takeo.sasai@ntt.com; yukinobu.nakajima@ntt.com; minami.takahashi@ntt.com; runa.kaneko@ntt.com; msnr.nakamura@ntt.com; asuka.matsushita@ntt.com; fukutaro.hamaoka@ntt.com; etsushi.yamazaki@ntt.com.)

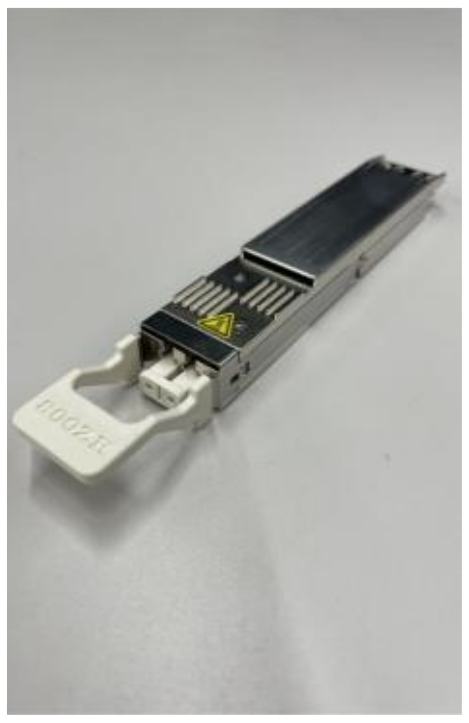

Fig. 1 OSFP module with LPM-integrated coherent DSP inside.

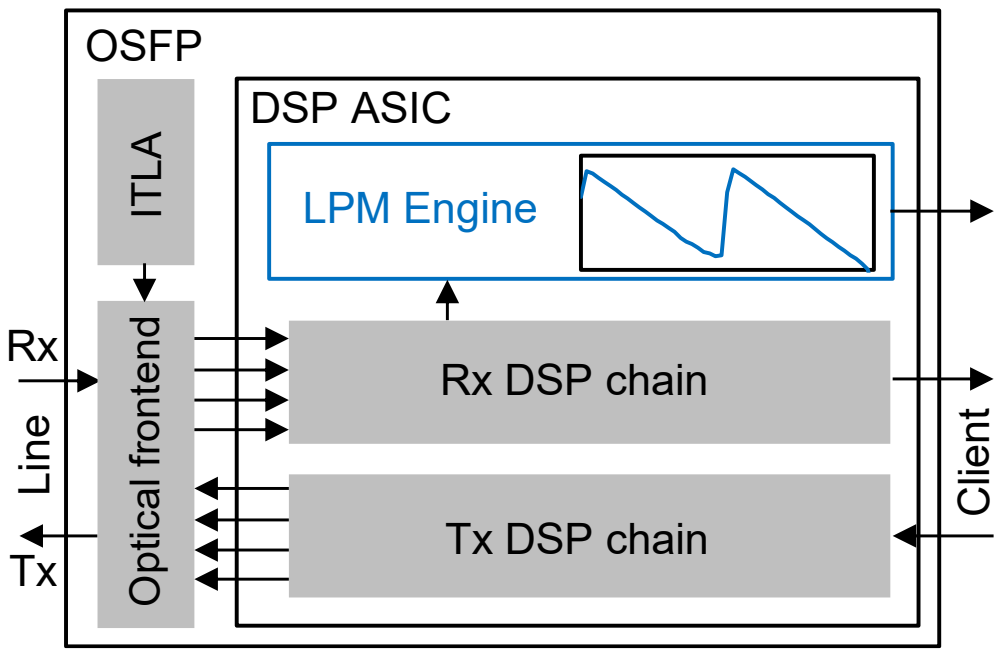

Fig. 2 Block diagram of OSFP module and coherent DSP ASIC, in which LPM engine is integrated alongside primary DSP chains.

nor probe light, and operates in-service without interrupting traffic. Being receiver-side, it also remains applicable to multi-span amplified links, where the measurement reach of backscattering-based fiber sensing is constrained by the optical isolators in in-line amplifiers.

Since its initial experimental demonstration in 2019 [11], LPM has overcome multiple barriers toward practical deployment. The major obstacles were: (i) insufficient estimation accuracy due to the absence of the optimal estimator; (ii) the lack of design guidelines; (iii) the need for high fiber launch power; and (iv) DSP implementation. For (i), a unified understanding of the various LPM algorithms was presented in [12], and the minimum variance unbiased estimator was derived in [13][14]. Regarding (ii), systematic methods for selecting the design parameters of LPM were established in [15], with related works reporting on noise analysis [16] and the appropriate choice of spatial sampling resolution [15][17]. For (iii), a number of approaches have been proposed to improve precision at lower launch power, such as L1/L2 regularization [18][19][20][21][22] and randomization of signal pattern [23]. As a result, LPM has recently become operable under the low-launch-power conditions as demonstrated using field fibers [47] and live networks [50]. Regarding (iv), implementations as software prototypes [24] and on a hardware accelerator [25] have been reported, yet integration into a coherent DSP ASIC chip has remained unrealized.

This paper focuses on (iv), particularly on the design of implementing LPM into a coherent DSP ASIC chip. We first review the principle of LPM and discuss the motivations, challenges, and enabling algorithm specific to its DSP implementation. We then present the on-chip LPM engine architecture and the key design considerations for achieving precise, in-service monitoring on a DSP. On this basis, we implement LPM on a coherent DSP chip—a 3-nm 800G low-power coherent DSP ASIC housed in an OSFP pluggable module (Fig. 1). This LPM engine performs all LPM computations, including power-profile averaging, entirely on chip (Fig. 2). We experimentally validate the on-chip LPM over links of up to 1005 km with a 2.1-km sampling resolution, successfully localizing multiple power anomalies along the link solely from standard ZR+ traffic received by the compact pluggable optics at the link endpoint [26]. We confirm that the LPM introduces no measurable impact on transmission quality or power consumption of the pluggable transceiver, demonstrating its compatibility with the stringent transmission performance requirements and power constraints of pluggable modules. Finally, we show that the on-chip LPM operates even on signals transmitted from third-party transceivers, in both 800ZR+ (16QAM) and 400ZR+ (QPSK) modes, confirming vendor- and modulation-format-independent operation.

## II. Distributed Fiber Monitoring on Coherent DSP

### A. *Longitudinal Power Monitoring*

The fundamental principle of LPM lies in the interaction between chromatic dispersion (CD) and fiber Kerr nonlinearity during signal propagation in fibers. LPM measures the distributed optical power by digitally extracting the Kerr nonlinearity, usually self-channel interference or self-phase modulation, from the received signal. As the optical signal travels through the fiber link, local nonlinear interference (NLI) is generated continuously at each position along the fiber due to Kerr effect, with each local NLI proportional to optical signal power (in weak nonlinear regime, where modern optical networks operate). Although what is observed at the receiver is a sum of the local NLIs, these local NLIs do not accumulate in a uniform or indistinguishable manner. This is because the CD continuously reshapes the original signal waveform during propagation, forcing the local NLI generated from the original signal to also carry a *position-dependent signature*. This signature allows spatial resolution. At the receiver, LPM exploits this position-dependent property to decompose the received total NLI into the local NLIs. A very rough but useful analogy is the orthogonal transformation such as Fourier transform (FT). In Fourier analysis, a waveform can be represented as a superposition of sinusoidal components, and the amplitude of each frequency component can be recovered through a linear transformation based on orthogonality among sinusoidal waves. Similarly, in LPM, total NLI observed at a receiver is a sum of many local NLIs, and the amplitude of each local NLIs, which is proportional to optical power, can be recovered through an appropriate linear transformation, although these local NLI components are not completely orthogonal to one another. Owing to these principles, LPM operates in-service on live traffic without dedicated probing hardware and remains applicable to multi-span amplified links. Detailed theoretical analysis can be found in [12][15], which underpin the LPM design discussed in this paper.

### B. *Motivations for DSP Implementation*

Beyond the features described above, the DSP integration of LPM is further motivated by the deployment trend of coherent pluggable transceivers. Pluggable transceivers, initially targeted at DCI with 400ZR [3], are increasingly deployed in metro and even long-haul links, as exemplified by ZR+ [4][5]. Embedding LPM in the coherent DSP ASIC would therefore allow the mass deployable pluggables to act not only as cost-efficient transmission elements but also as distributed monitoring points across diverse optical network infrastructures from DCI to long-haul [2], enabling ubiquitous network monitoring. However, due to the computational and memory requirements discussed later, implementing LPM on a DSP ASIC has been considered highly challenging.

The dependence on external processing indeed introduces multiple barriers to widespread deployment of LPM in networks. First, the signal waveform extraction to external server is bottlenecked by interface bandwidth and requires implementation support for exporting large waveforms across the DSP, pluggable module, host switch, and external controller. Moreover, since the available interfaces, data format, and access methods depend on the system and vendor, the processing pipeline must be individually developed and integrated for each deployment. Finally, exporting raw waveform data outside the transceiver poses a risk of exposing potentially sensitive signal information [27]. As a result, implementing LPM within the DSP is highly desirable for practical and widespread adoption.

## III. Algorithm Design

### *A. Best estimator*

The principle described in the previous section can be formulated as a linear least squares estimation problem [13]. Let $\boldsymbol{A_1}$ denote the time vector of the total NLI observed at the receiver, and our estimation target $\gamma'(z) = \gamma(z)P(z)$ be the product of the fiber nonlinearity coefficient and the optical power at position $z$. As described above, the total NLI is the accumulation of the local NLIs $\gamma'(z)\boldsymbol{g}(z)$ over the link length (one can see that the local NLIs are proportional to the local optical powers), and thus can be expressed as $\boldsymbol{A_1} = \int \gamma'(z)\boldsymbol{g}(z)dz$. Discretizing this integral yields a matrix-vector form:

$$\boldsymbol{A_1} = \boldsymbol{G}\boldsymbol{\gamma}' \tag{1}$$

where $\boldsymbol{G}$ is a matrix representing nonlinear fiber propagation and is determined by the group velocity dispersion of the fiber, signal symbol rate, modulation format, and pulse shaping. Its matrix size is the time sample size $N_t$ times spatial points $N_z$. Based on (1), the least-squares estimator for spatial power profile is obtained as [13]:

$$\widehat{\boldsymbol{\gamma}}' = (\mathrm{Re}[\boldsymbol{G}^\dagger\boldsymbol{G}])^{-1}\mathrm{Re}[\boldsymbol{G}^\dagger\boldsymbol{A_1}] \tag{2}$$

This estimator is the minimum variance unbiased estimator (MVUE) when the received noise is additive white Gaussian. A regularization term can be incorporated into (2) to mitigate ill-conditioning and suppress noise amplification, at the cost of introducing estimation bias, as discussed in [18][21]. Extension to dual-polarization signals via the Manakov model is detailed in [13][28].

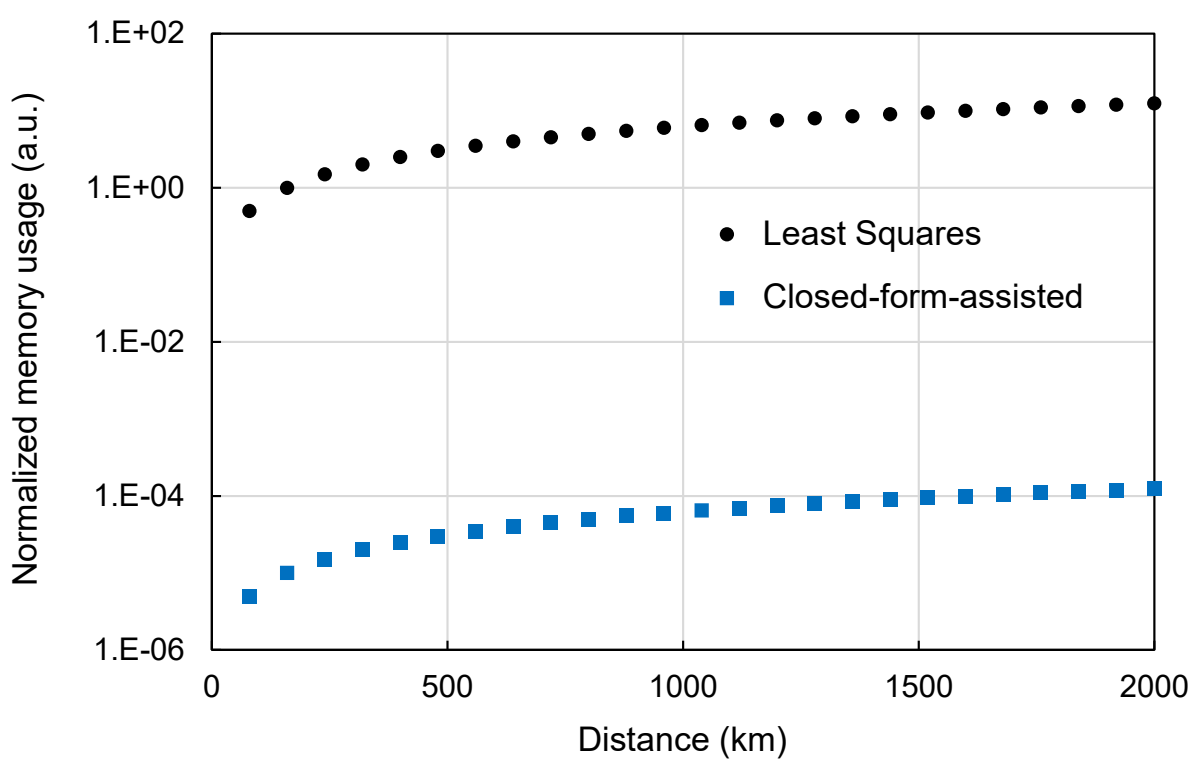

Fig. 3 Memory usage of least-squares-based and closed-form-assisted LPM, normalized to the value at 160 km.

### *B. Challenges for DSP Implementation*

Implementation of the least squares estimator (2) faces practical limitations arising from large memory consumption. For example, a typical configuration of LPM may use discretization with $N_z$~1000 positions (e.g., 1 km resolution for 1000 km links) and $N_t$~ $10^5$ time samples for matrix $\boldsymbol{G}$, necessitating memory allocation on the order of hundreds of megabytes to gigabytes, which is incompatible with power and area constraints of DSP ASIC, particularly for pluggable applications. Furthermore, in (2), the matrix inversion is required, also leading to substantial computational complexity and memory usage that are not well suited to DSP implementation. Consequently, most prior demonstrations of LPM have relied on offline processing outside transceivers.

Several memory-efficient approaches were proposed in [29][30], based on the observation that the $\boldsymbol{G}^\dagger\boldsymbol{G}$ in (2) becomes a Toeplitz matrix [12][15] (i.e., linear convolution), under the assumption that the signal follows a stationary complex Gaussian process everywhere along a fiber link under test. In this case, its inverse can be approximated as a deconvolution (though this approximation requires careful mathematical consideration; see [31][15]), thereby avoiding the need to store and invert matrices.

However, a critical factor to consider is the modulation-format dependency of LPM. It is well-known that fiber nonlinearity strongly depends on signal modulation format and its amplitude distribution, particularly at locations near the transmitter side [32][33]. Since LPM relies on fiber nonlinearity to estimate optical power, it also shows dependency on modulation format, as pointed out in [12]. Consequently, the memory-efficient deconvolution LPM such as [29][30], which assumes the transmitted signal to be Gaussian, cannot be directly applied as it suffers from significant degradation of LPM accuracy when practical modulation formats such as QPSK are used (see Fig. 4 and Fig. 5). State-of-the-art commercial coherent DSP ASICs support various transmission modes including QPSK, 16QAM, and PCS-QAMs with various shaping distributions. Conventional matrix-based approaches such as LS-LPM in (2) therefore construct $\boldsymbol{G}$ or $\boldsymbol{G}^\dagger\boldsymbol{G}$ from signal data for each modulation format to account for format-dependent nonlinearity and to obtain accurate LPM results, which incurs prohibitive memory and computational costs for on-chip implementation. In [34], authors proposed a multi-stage method that partially computes LPM at the receiver while calculating the memory-hungry $\boldsymbol{G}^{\dagger}\boldsymbol{G}$ and its matrix inversion beforehand using offline processing, still requiring multiple large matrices to be stored for each transmission mode. In [35], to avoid matrix inversion, the block least mean squares algorithm was suggested, though this approach still demands large sample storing for long-haul applications.

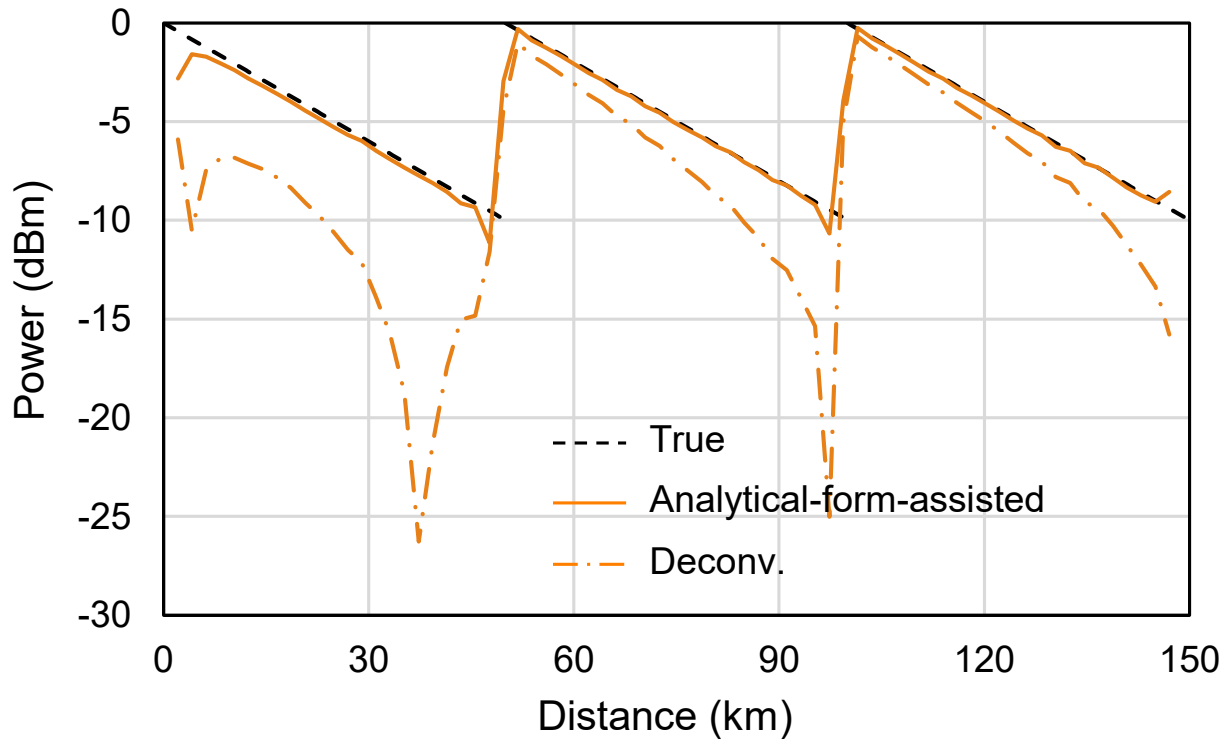

Fig. 4 Numerical results of deconvolution-based and analytical-form-assisted LPM for QPSK.

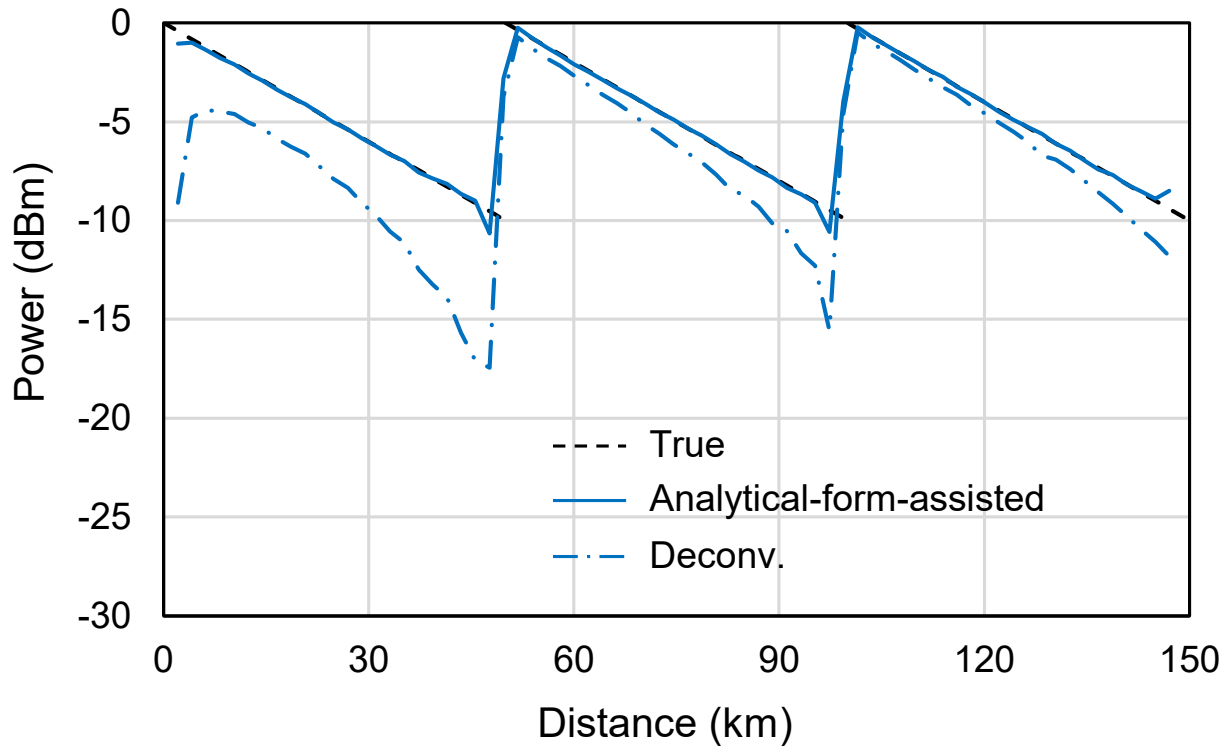

Fig. 5 Numerical results of deconvolution-based and analytical-form-assisted LPM for 16QAM.

### C. *Analytical-form-assisted Estimator*

To address the memory and complexity issues, an analytical-form-assisted LPM algorithm is beneficial. The starting point is the structure of the matrix to be inverted in (2). Under the Gaussian signal assumption, $\boldsymbol{G}^{\dagger}\boldsymbol{G}$ becomes a Toeplitz (i.e., linear convolution) matrix whose entries depend only on the distance between two positions [15]:

$$(\mathrm{Re}[\boldsymbol{G}^{\dagger}\boldsymbol{G}])_{ij} \propto h_R(z_j - z_i) \tag{3}$$

where $h_R(z)$ is the real part of the spatial correlation function (SCF), which represents the correlation between the nonlinear waveforms generated at two positions separated by $z$. In [36], the Fourier transform of the SCF for Nyquist-limit signals was derived in closed form in the wavenumber domain as a function only of the group velocity dispersion and the symbol rate. With this expression, the matrix in (3) does not need to be constructed from signal data or stored, since it is fully specified by a few link and signal parameters, which reduces the memory consumption by 99.999% as shown in Fig. 3. Following this approach, an analytical formulation of the least-squares computation is adopted that accommodates any modulation format, including QPSK, 16QAM, and PCS-QAMs, without explicitly constructing the signal-dependent matrices, enabling memory-efficient and format-agnostic LPM on the DSP ASIC. This formulation statistically preserves the LS-LPM solution exactly so as not to sacrifice the accuracy of LPM. This exactness is important because, in commercial transceivers, noise and distortion are dominant, and LPM must extract weak fiber nonlinearity from received signals among these impairments [37]. An algorithmic approximation would further degrade the already limited SNR of the monitoring observable, making it essential to preserve the statistical optimality of the LS-LPM solution. Fig. 4 and Fig. 5 numerically compare the LPM accuracy of deconvolution and analytical-form-assisted approaches for QPSK and 16QAM signals, respectively. Transmission of a single-polarization 118.2-GBd signal over a 50 km × 3 span link was emulated using the split-step method with a fiber loss of 0.2 dB/km, a CD of 17 ps/nm/km, and nonlinear constant of 1.3 $\mathrm{W}^{-1}\mathrm{km}^{-1}$. The deconvolution method exhibits significant degradation for practical modulation formats because it assumes Gaussian transmitted signals. In contrast, analytical-form-assisted approach achieves high estimation fidelity regardless of the modulation format.

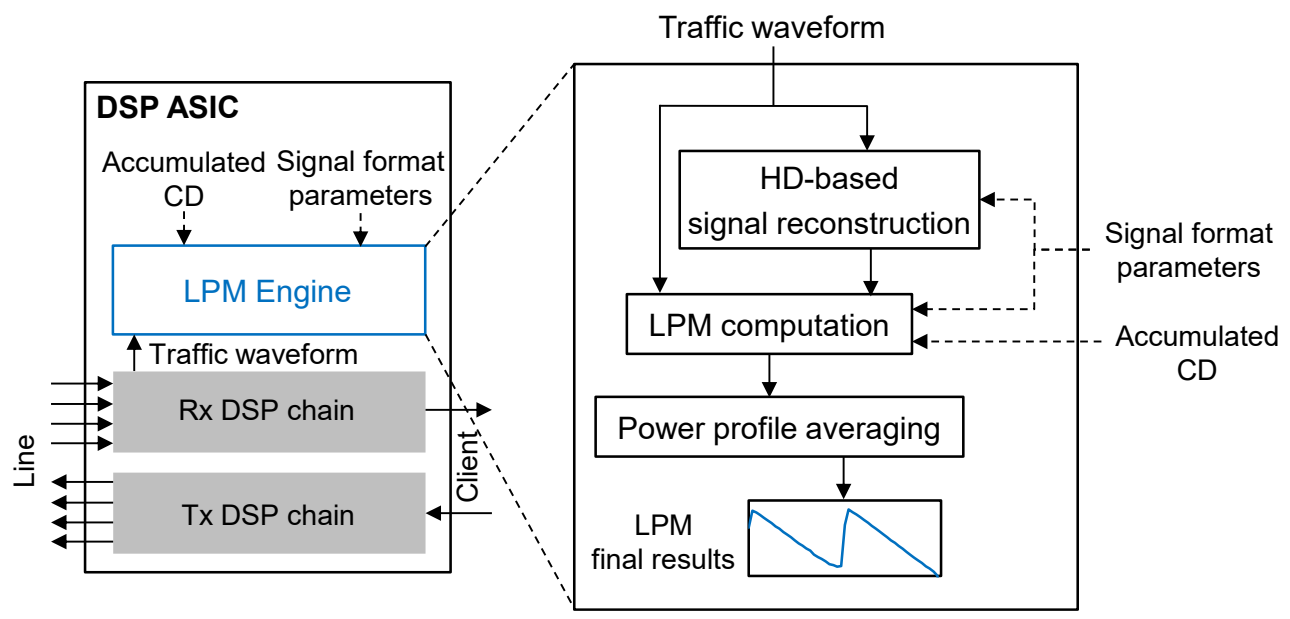

Fig. 6 Architecture of the LPM engine in coherent DSP ASIC.

## IV. DSP Implementation

### A. *LPM Engine Architecture*

In this work, the LPM is implemented within the coherent DSP ASIC chip. Fig. 6 shows the block diagram of the DSP and LPM engine architecture. The LPM engine operates as a parallel processing unit alongside the primary DSP chain, tapping received traffic waveforms from the Rx DSP pipeline at a stage where the linear impairments have been compensated. All computations for LPM are entirely done within Rx DSP chip. The LPM engine requires three inputs: the received waveforms, accumulated CD, and signal format parameters, all of which are readily available within the receiver DSP. The accumulated CD is usually estimated during demodulation as a part of standard DSP processing. The signal format parameters, namely the shaping distribution and symbol rate, are predetermined as transceiver operating parameters. Although the transmitted waveforms are required for LPM computation in theory, they are not assumed to be known a priori as it is reconstructed from the received waveforms within the LPM engine. Consequently, the LPM results are obtained in a fully blind manner without external, additional inputs.

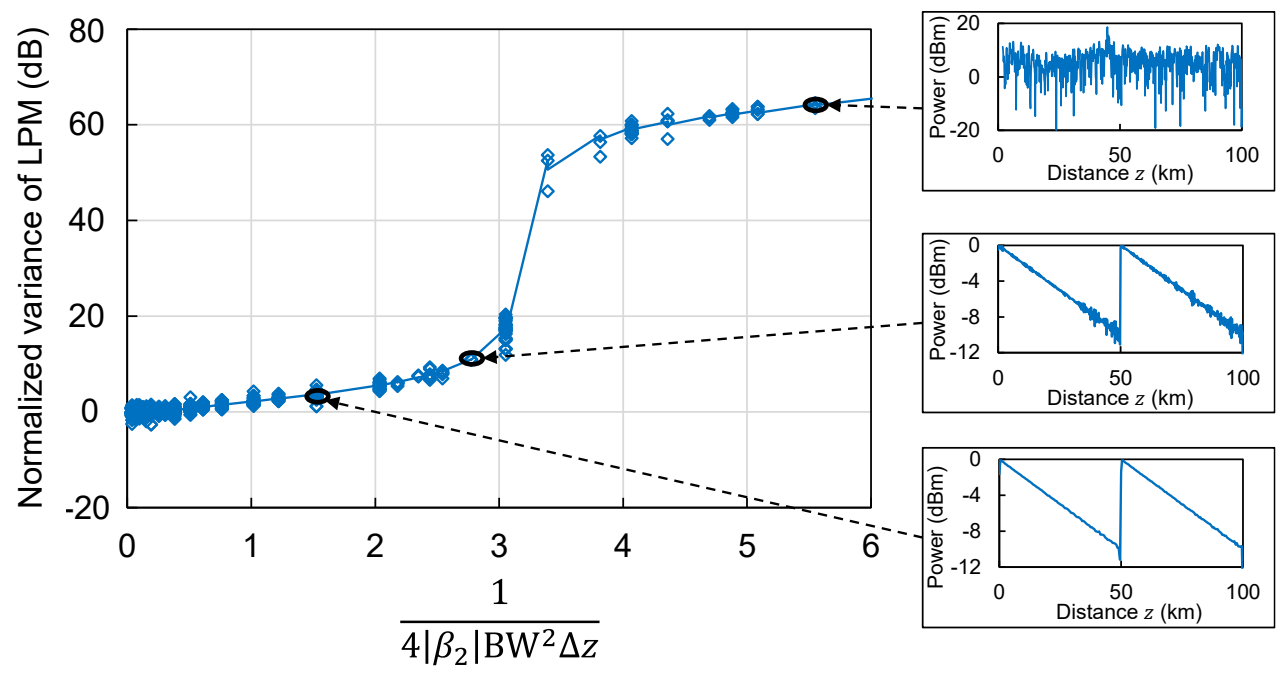

Fig. 7 Noise variance of LPM as a function of reciprocal of CD step. LPM noise increases sharply beyond the threshold $\pi$. The plots were obtained by simulations and represent all possible combinations of the symbol rate BW $\in \{64, 128, 256\}$ GBd, CD coefficient $\beta_2 \in \{-5, -10, \dots, -30\}$ ps$^2$/km, link length $L \in \{100, 250, 500\}$ km, and $\Delta z \in \{0.5, 1, 2, 4\}$ km. The received noise corresponding to an SNR of 17 dB was added. For simplicity, single-polarization Gaussian-distributed signals with Nyquist spectra were used.

The LPM engine consists of three blocks: reconstruction of transmitted signals, analytical-form-assisted LPM core computation, and profile averaging. For the signal reconstruction, the hard decision (HD) is used to reconstruct transmitted signals, eliminating the need for dedicated probing signals or special training sequences. For LPM core computation, the analytical-form-assisted computation is used, suppressing the hardware overhead associated with the LPM implementation. The core logic area occupies less than 1% of the total DSP chip area, imposing negligible power consumption and hardware overhead and thus complying with strict size- and power-constraints of the OSFP. For profile averaging, successive profile estimates are accumulated and averaged on chip, which improves the power-profile SNR and eliminates the need to export intermediate results off the module. The LPM engine does not need to process the entire traffic; it captures blocks of received symbols intermittently, and the total number of symbols is a design parameter chosen to meet the required power-profile SNR following the guideline in [15]. The final LPM results are retrieved via the I$^2$C interface of the OSFP module. LPM can be enabled or disabled without affecting the operation of the main DSP chain.

### *B. Hard Decision for In-service Monitoring*

LPM in principle requires the transmitted signal as it is an inverse boundary-value problem, where the distributed link parameters are estimated from the boundary observations, i.e., the transmitted and received signals. Indeed, the least squares algorithm in (2) uses the Tx signal, for example, to generate $\boldsymbol{G}$ matrix. Several prior demonstrations therefore assumed the transmitted signal to be known rather than operating on in-service traffic [10]. In this work, the transmitted signal is reconstructed from the received signal within the Rx DSP chip by applying hard decision (HD). Consequently, the LPM engine operates directly on in-service payload traffic without special probing or training sequence, enabling communication and forward distributed monitoring simultaneously. Notably, this approach also allows LPM to operate using signals transmitted from third-party transceiver signals, which is critical for multi-vendor interoperability in modern optical networks, where transceivers from different manufacturers coexist. However, the HD inevitably introduces decision errors, which give rise to two types of error. The former is a systematic error in which a downward offset appears in the estimated power profile [38]. Several compensation methods for this offset have been proposed in [39][40]. The latter is a statistical error in which the power profile becomes noisy and the precision degrades, appearing as a penalty in the power-profile SNR [39]. These penalties are more pronounced when using high-order modulation formats, such as 64QAM, under the same amount of noise. To mitigate the HD-induced penalty, when pilot symbols (PS) or training sequences (TS) are available in the DSP frame as defined, for instance, in the ZR implementation agreement [3], these known symbols are used in place of HD at their positions in the reconstructed transmitted signal, while the payload symbols are reconstructed by HD. Another approach based only on PS and TS to mitigate the HD penalty has also been proposed in [41].

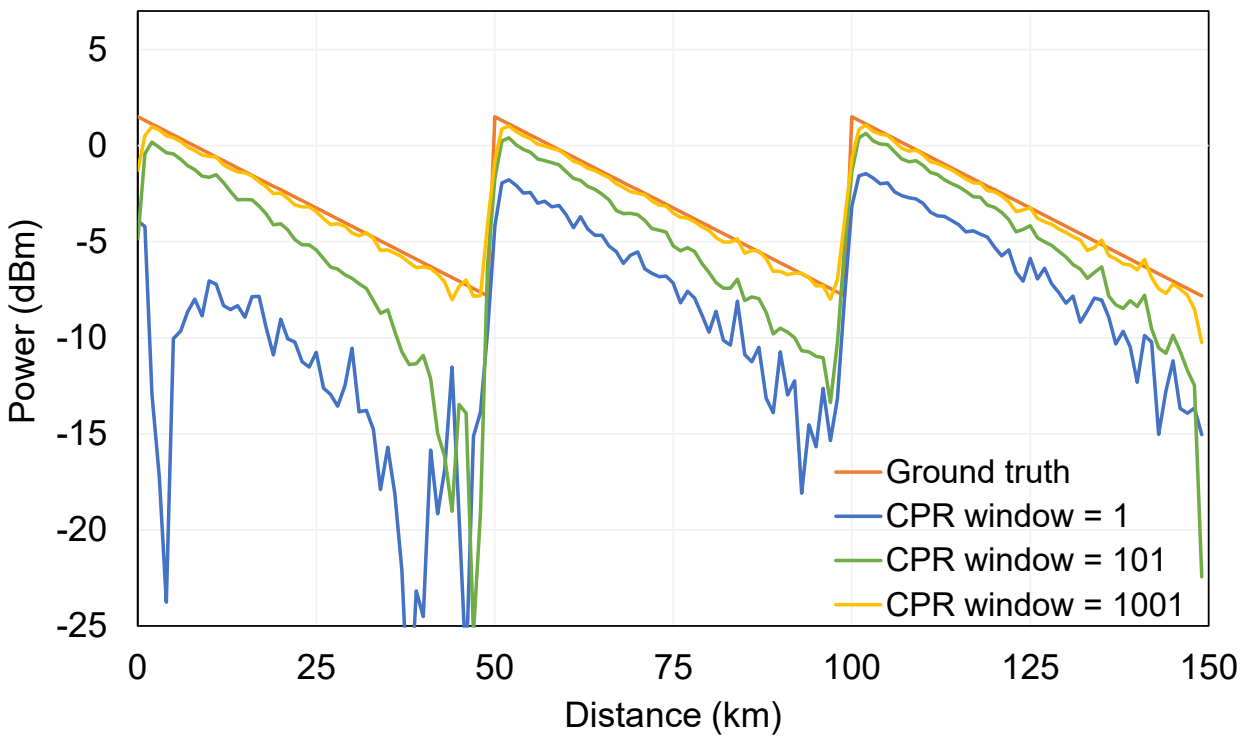

Fig. 8 Numerical results of LPM for different CPR averaging window lengths with a laser linewidth of 100 kHz. A 100-GBd PCS-64QAM signal was used, and an amplifier noise figure of 5 dB was assumed.

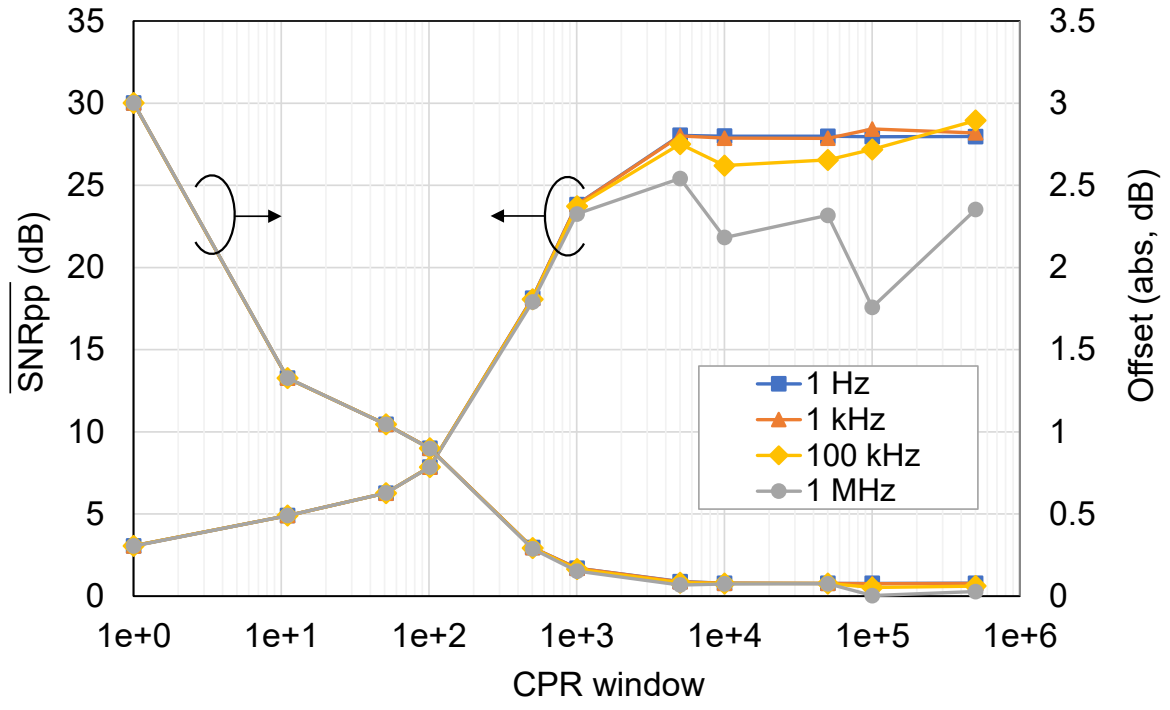

Fig. 9 Power-profile SNR and systematic offset as a function of CPR averaging window length for various laser linewidths. Conditions are the same as in Fig. 8.

In principle, an HD-error-free implementation is possible by reconstructing the transmitted signal from the corrected bits at the FEC output. This would require re-encoding, symbol-mapping, and PS/TS insertion, incurring substantial computation and memory.

Table I Relevant LPM demonstrations

| | Implementation location | Receiver type | Traffic-signal-based monitoring[1] | Interop. demo. | Distance |
|---|---|---|---|---|---|
| [11] | Offline | Discrete | No | No | 260 km |
| [46] | Offline | Discrete | No | No | > 10000 km |
| [38] | Offline | Transponder | Yes | No | 900 km |
| [47] | Offline | Transponder | No | No | 163 km |
| [48] | Offline | Transceiver | Yes | No | 200 km |
| [49] | Rx-adjacent + controller | Transponder | Yes | No | ~400 km |
| This work | Fully on DSP chip | Pluggable (OSFP) | Yes | Yes | 1005 km (400ZR+) 450 km (800ZR+) |

[1] Yes: uses payload data (i.e., Tx symbols are unknown)

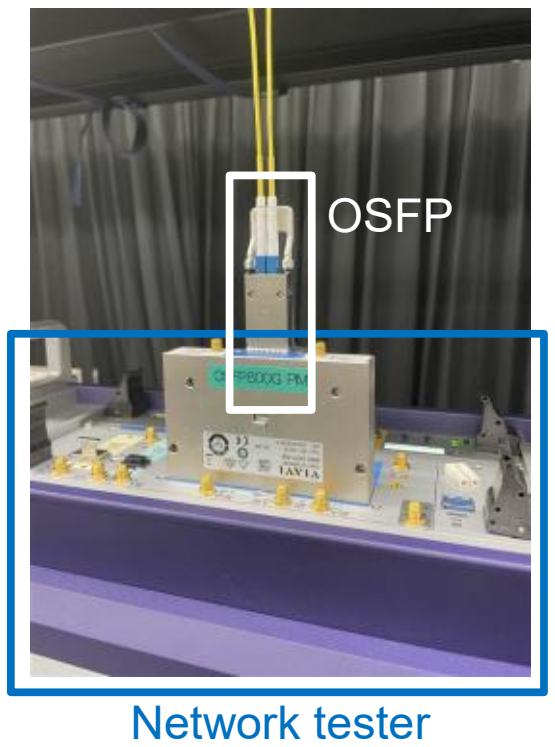

Fig. 10 Tx-side pluggable transceiver hosted on a commercial network tester generating client traffic.

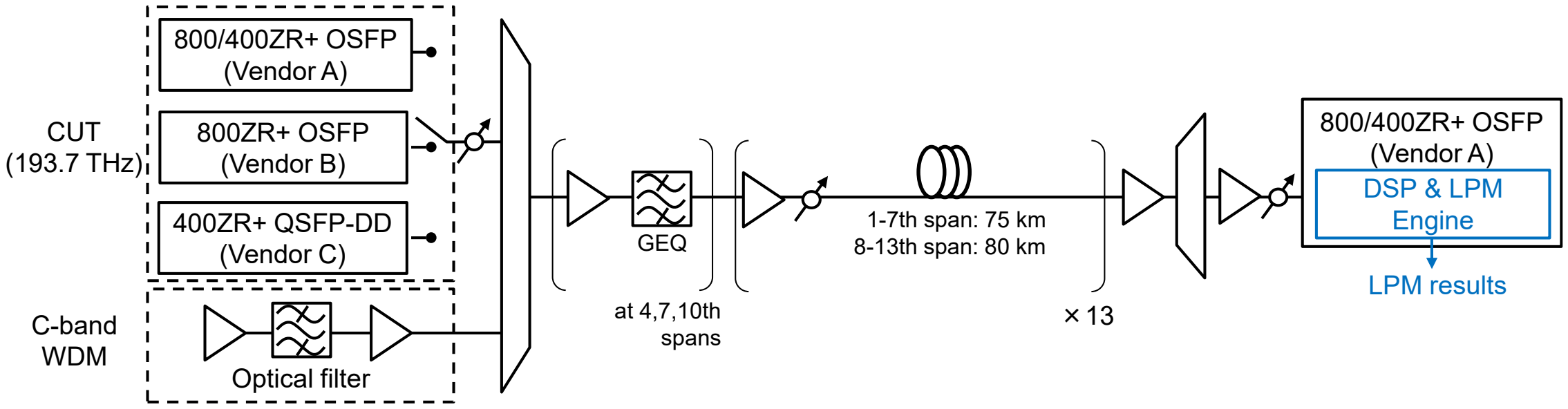

Fig. 11 Experimental setup for 1005-km transmission.

## C. *Selection of Spatial Sampling Resolution*

Another design consideration is the selection of the spatial sampling resolution $\Delta z$. Unlike backscattering-based sensing, the spatial sampling resolution of LPM is not set by a physical condition such as pulse width. Instead, it is a design parameter chosen in the DSP as a trade-off among the resolution, estimation precision, and slope bias in the estimated power profiles. In the least squares LPM, selecting an excessively fine resolution results in the degradation of the power-profile SNR, as pointed out in [13] [17]. The sampling resolution $\Delta z$ should therefore be chosen large enough to suppress the LPM noise, yet fine enough to preserve the spatial resolution and estimation fidelity required by the intended applications [15]. This trade-off stems from the fact that the spatial resolving mechanism of LPM is based on CD-induced reshaping of the signal waveform during propagation, so two positions are separable only when the waveform changes appreciably between them. If $\Delta z$ is set too small, the waveform barely changes from one step to the next, so adjacent positions cannot be told apart and their resulting power estimates become noisy (i.e., the problem is ill-conditioned). Conversely, an overly coarse $\Delta z$ degrades localization accuracy and introduces a slope bias in the estimated power profiles [13][17]. Fig. 7 illustrates how the estimation variance of LPM (i.e., noise power in LPM) depends on the sampling resolution $\Delta z$ [15][36]. The variance remains suppressed for a large $\Delta z$ region (left side of the figure), but rises sharply once $\Delta z$ is reduced below a threshold determined by the chromatic dispersion and the signal bandwidth. Beyond this threshold (~3.1 on the horizontal axis, shown to be exactly $\pi$ in [36]), the estimation problem becomes ill-conditioned and the profile rapidly degrades. In this work, $\Delta z$ was set to 2.1 km, which lies within this stable region while still resolving km-scale loss events. The total number of symbols used for one profile was on the order of $10^6$ to $10^7$.

## D. *Impact of Carrier Phase Recovery*

The carrier phase recovery (CPR) also affects the performance of LPM. The CPR is intended to track and compensate for the laser phase noise, but it inevitably absorbs part of the Kerr nonlinear phase as well [42][43]. Since this nonlinear phase is the quantity used by LPM for estimation, excessive phase tracking by CPR degrades its estimation accuracy [44]. Fig. 8 shows numerical evaluation of the estimated power profiles obtained with different CPR averaging window size under a 100-kHz laser linewidth. As the window increases, the estimated profile converges toward the ground truth, indicating that the suppression of excessive phase tracking allows restoring the nonlinear phase used by LPM. Fig. 9 further plots the power-profile SNR (left axis) and systematic offset (right axis) as a function of the CPR averaging window size, confirming that the power-profile SNR improves and the offset vanishes as the window increases. Here, the systematic offset is defined as the mean difference in dB between the estimated and the true power over positions excluding span boundaries, while the power-profile SNR is defined, following [15][44], as $\overline{\mathrm{SNR}_{\mathrm{pp}}} = \frac{1}{N_z}\sum_m \frac{{\gamma'_m}^2}{V\left[\widehat{\gamma'_m}\right]}$, where $\gamma'_m$ and $V\left[\widehat{\gamma'_m}\right]$ are the true value and the estimation variance at position $z_m$, respectively. A larger window, however, degrades the ability of CPR to track the laser phase noise itself, and the resulting residual phase noise penalizes the LPM accuracy and communication signal quality. Therefore, the window length should be selected so as to achieve sufficient laser phase noise tracking while avoiding excessive cancellation of nonlinear phase. Fig. 9 provides a quantitative basis for this selection, and with an appropriate window, both LPM accuracy and communication signal quality can be maintained even for practical laser linewidths, as demonstrated with pluggable transceivers in this work.

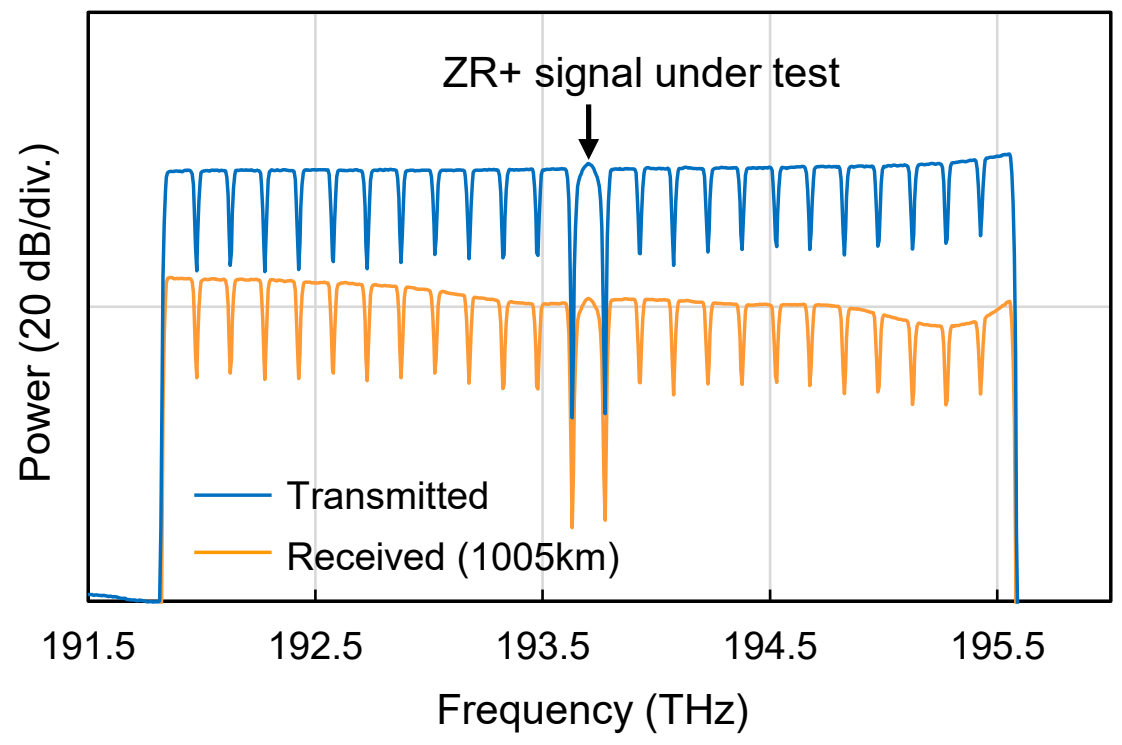

Fig. 12 Transmitted and received (1005km) WDM spectra.

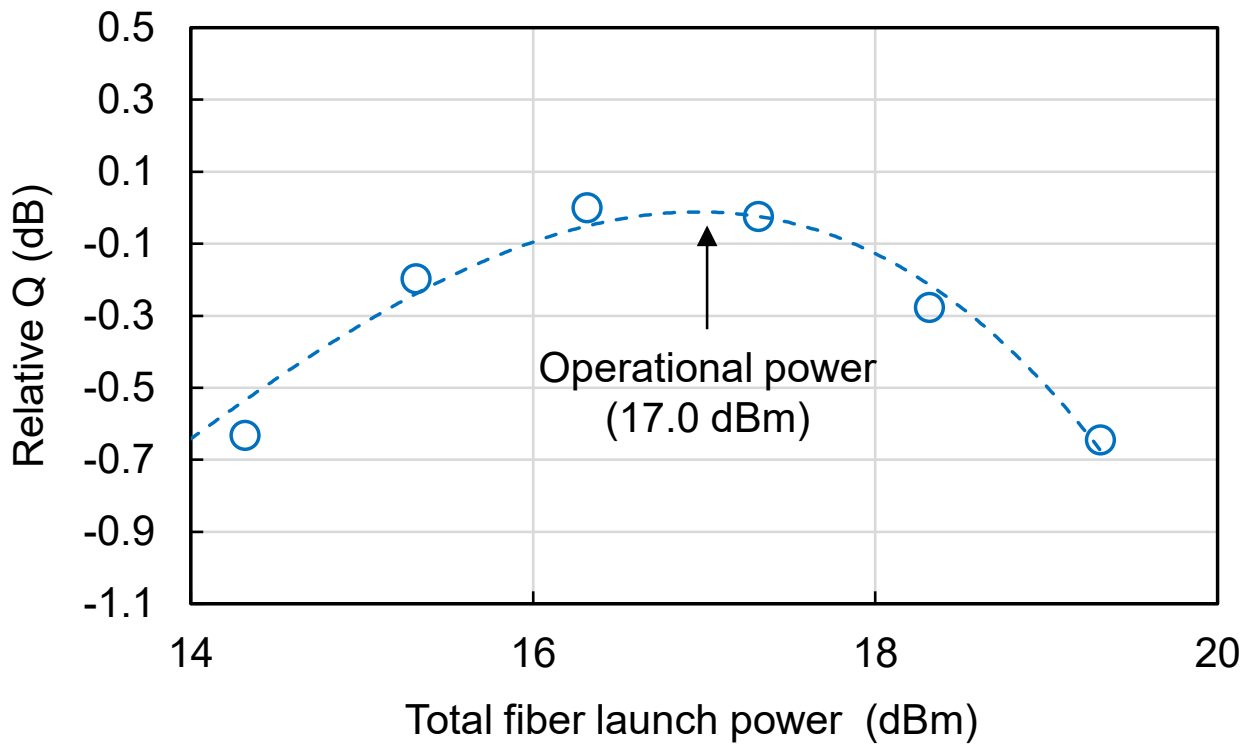

Fig. 13 Relative Q value versus fiber launch power after 1005-km transmission. The LPM is operated at the launch power that maximizes the Q factor.

## V. Experimental Validation

In this section, we report the first experimental demonstration of LPM on a coherent DSP ASIC (Table I). Based on the design considerations presented in the preceding sections, we implement the LPM on a 3-nm 800G DSP ASIC chip. This LPM-integrated DSP is then housed in an OSFP transceiver module and validated over distances of up to 1005 km. The on-chip LPM operates by processing only received ZR+ traffic, without a dedicated probe or training sequence. We further confirm that the on-chip LPM incurs no measurable impact on either Q-factor or module power consumption, and that it remains effective even when receiving signals from third-party transmitters, demonstrating multi-vendor interoperability.

### *A. Experimental Setup*

Fig. 11 shows the experimental setup. The transmitter employed three different pluggable modules to enable multi-vendor validation. One is an OSFP module supplied by Vendor A with our DSP inside, supporting both 800ZR+ (DP-16QAM) and 400ZR+ (DP-QPSK) at the same 118.2 GBd. Second is an OSFP from Vendor B with a different DSP vendor, supporting 800ZR+. Third is a QSFP-DD module from Vendor C with another distinct DSP vendor, supporting 400ZR+. These pluggable modules were hosted on a VIAVI ONT-804 network tester with client traffic as shown in Fig. 10. As such, no specialized signal was used and the LPM operated on standard ZR+ traffic signals.

The optical link configuration consisted of two test scenarios: 1005 km using 400ZR+ (75 km × 7 spans + 80 km × 6 spans) and 450 km using 800ZR+ (75 km × 6 spans). All fibers are compliant with G.652D with an average attenuation of 0.186 dB/km and chromatic dispersion of 16.8 ps/nm/km. Although the experiment was conducted over G.652D fibers, LPM has been shown to maintain consistent accuracy across various fiber types, including G.654-compliant pure-silica-core fibers, at their respective optimal launch powers [45]. As shown in Fig. 12, the WDM transmission was emulated using spectrally shaped amplified spontaneous emission noise, creating 24 WDM channels on a 150-GHz grid spacing surrounding the channel under test (CUT) at 193.7 THz and resulting in a total bandwidth of 3.75 THz. Note that LPM does not require any information on the WDM channel loading, since the XPM from the co-propagating channels is uncorrelated with the channel under test and acts as additional stochastic noise in the estimation. The total optical power launched into the fiber was set to 17.0 dBm, optimized to maximize the Q-factor of CUT (Fig. 13). Gain equalizers (GEQs) were inserted after the 4th, 7th, and 10th spans to compensate for the accumulated gain tilt of the EDFA chain. Loss anomalies were inserted using variable optical attenuators at 175 km and 724 km (each located in the middle of spans) and an EDFA gain anomaly was introduced at 450 km (the input of the 7th span) to demonstrate LPM's localization capability.

The receiver was another OSFP module from Vendor A with LPM-integrated DSP. Again, the DSP in the receiving OSFP is supplied by a different DSP vendor than those used in Vendor B and Vendor C at the Tx side, enabling LPM's multi-vendor interoperability verification. The LPM spatial sampling resolution was set to $\Delta z$ = 2.1 km. In the LPM engine, the transmitted signal was reconstructed by HD for the payload and by the known symbols in the PS and TS positions in the ZR frame. Throughout measurements, the post-FEC bit-error rate remained error-free during LPM operation, since the LPM engine only taps the received signal already processed by the coherent DSP and therefore does not perturb the communication, enabling simultaneous communication and distributed fiber monitoring. OTDR references were obtained independently to enable quantitative comparison with LPM results. Note that the OTDR traces are shifted by a constant offset determined in the least-squares sense with respect to the LPM results.

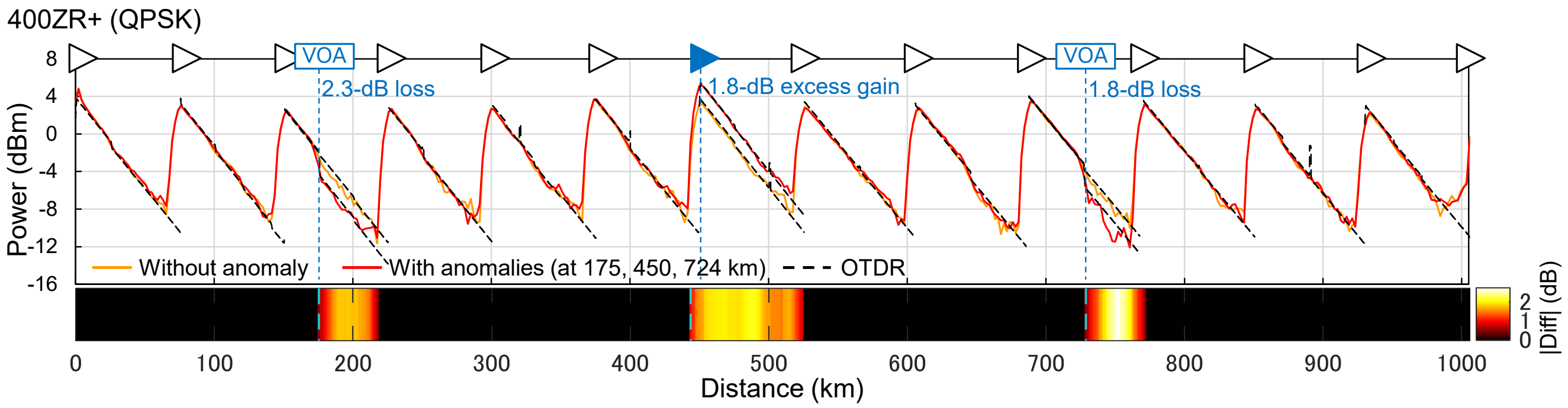

Fig. 14 On-chip LPM results over 1005 km using 400ZR+ (QPSK) with and without power anomalies at 175, 450, and 724 km.

## B. *On-chip LPM Results over 1005 km*

Fig. 14 shows the on-chip LPM results over 1005 km using the 400ZR+ (QPSK) signal under both normal and anomalous conditions. Under normal operating conditions, the power profiles clearly captured both the fiber attenuation slopes and the EDFA gain steps along the link and showed agreement with the OTDR reference with an RMS error of 0.43 dB excluding the inter-span transition regions. Even when three anomalies were introduced in the link, highly stable and reproducible power profiles are observed in spans without anomalies. Meanwhile, in the anomalous spans, clear power deviations are observed, demonstrating localization capability. The heatmap below shows the normal-anomalous differences that exceed 4σ of the LPM noise, clearly localizing all three impairments along the link. Fig. 15 shows the on-chip LPM results for the second scenario over 450 km link using 800ZR+ (16QAM). Comparison with OTDR measurements yielded an RMS error of 0.60 dB, comparable to the 0.43 dB obtained with QPSK over 1005 km. This consistency demonstrates that the on-chip LPM is less dependent on the modulation format. Although only uniform QAM formats were tested here, the analytical-form-assisted estimator also accommodates PCS-QAM by design, supporting its applicability across different transceiver modes.

## C. *Transmission Quality and Power Consumption Stability*

Simultaneous operation of the LPM engine and data transmission was verified through continuous measurement of signal quality and power consumption metrics. Fig. 16 shows the pre-FEC Q-factor recorded continuously while the LPM was switched on and off. The variation remained within ±0.04 dB throughout the measurement, demonstrating that the LPM operation introduces no measurable degradation to signal quality. This is expected by design, since the LPM engine only taps the received signal from the main Rx DSP chain in the ASIC. Post-FEC BER also remained error-free in both operational states, confirming simultaneous distributed monitoring and communication without mutual interference. Fig. 17 shows the power consumption measurements of the OSFP module. No measurable increase in overall module power consumption was observed irrespective of LPM on/off operation, indicating that the on-chip LPM introduces negligible power consumption overhead and thus satisfies the strict power constraints of the OSFP module.

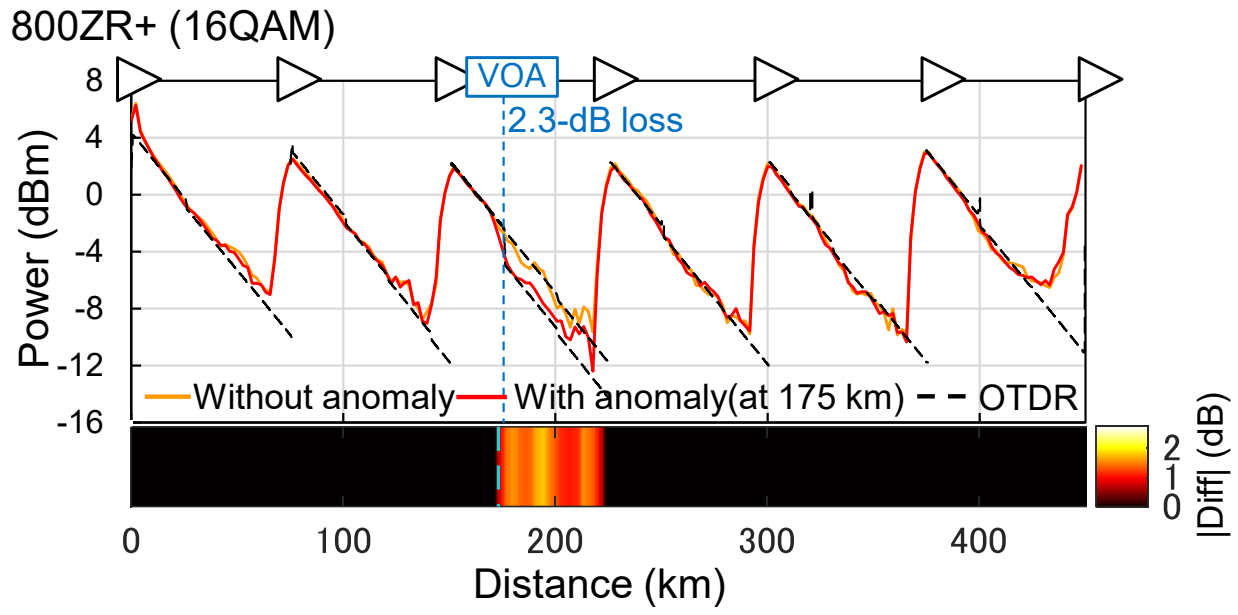

Fig. 15 On-chip LPM results over 450 km using 800ZR+ (16QAM) with and without power anomaly at 175 km.

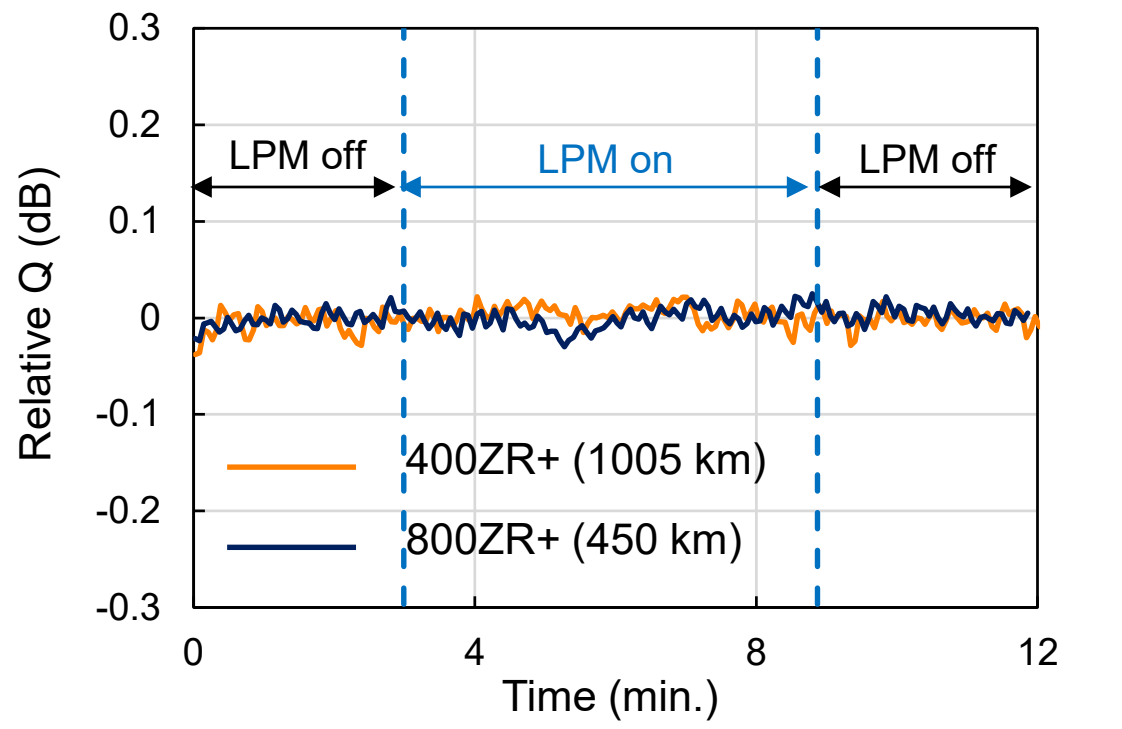

Fig. 16 Pre-FEC Q factor stability during on/off switching of LPM engine

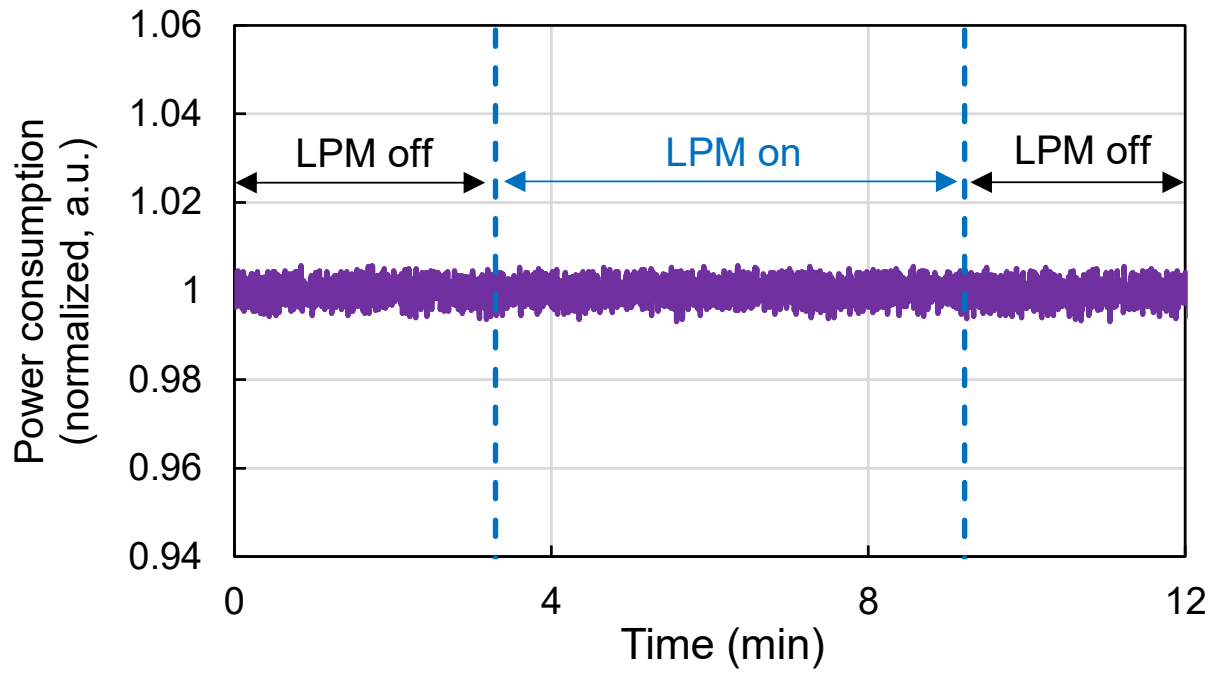

Fig. 17 Power consumption stability during on/off switching of LPM engine

## D. *Multi-Vendor Interoperability*

In this subsection, multi-vendor interoperability was validated across both modulation formats. In the 400ZR+ mode

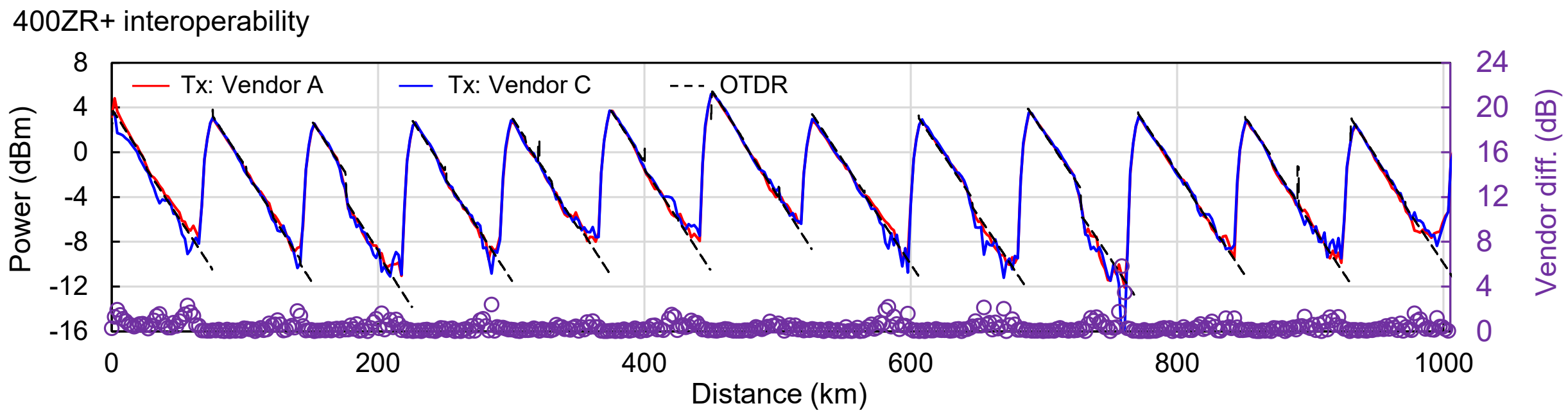

Fig. 18 Multi-vendor interoperability: on-chip LPM results in 400ZR+ (QPSK) mode with transmitter modules from different vendors. Circles represent the difference between the power profiles obtained with the two vendors' transmitters (right vertical axis).

(QPSK), transmitter modules from Vendor A (OSFP) and Vendor C (QSFP-DD) were each paired with the Vendor A receiver containing the LPM-integrated DSP, as shown in Fig. 18. The LPM results using both transmitters demonstrated agreement, with an RMS deviation between vendor profiles of 0.63 dB. This consistency confirms that the hard-decision-based transmitter signal reconstruction operates correctly across different transmitter DSP implementations, where transmitter waveform characteristics are not known a priori. A notable observation is that a relatively large discrepancy between Vendor A and Vendor C appears in the first span on the transmitter side. This is presumably because the waveform transmitted by Vendor C deviates from the reference waveform assumed by the LPM engine since transmitter bandwidth characteristics, other front-end imbalances, and even compensation policies differ across vendors. These waveform differences likely account for the power-profile deviation observed in the first span. As the signal propagates along the fiber, however, this discrepancy diminishes. We attribute this observation to the inter-symbol interference induced by CD, whose averaging effect among neighboring symbols gradually masks the transmitter-side waveform distortion. The 800ZR+ (16QAM) interoperability test employed transmitter modules from Vendor A and Vendor B (Fig. 19). The measured RMS deviation between the two vendor power profiles was 0.70 dB, comparable to the 400ZR+ (QPSK) case. These results demonstrate LPM's interoperability across different pluggable module and DSP vendors as designed in the previous section, a critical requirement for practical deployment in modern networks.

## VI. Conclusion

In this paper, we have reported the realization of the first coherent DSP ASIC capable of distributed fiber monitoring, achieved by implementing LPM on a 3-nm 800G low-power DSP housed in an OSFP pluggable module. We presented the key design considerations spanning the enabling algorithm, the architecture for in-service monitoring, parameter selection for retaining estimation fidelity, and the interplay with DSP compensation blocks. The on-chip LPM was validated over links of up to 1005 km using standard 400ZR+ and 800ZR+ traffic, localizing multiple power anomalies along the link. No measurable impact on Q factor or module power consumption was observed during LPM operation, satisfying the strict requirements of pluggable modules. We also demonstrated the LPM's multi-vendor interoperability, confirming its consistent operation using signals transmitted from third-party transceivers. These results establish coherent DSP and pluggable transceivers as spatially resolved monitoring elements of optical networks, extending their role beyond data transport toward detailed physical-layer visibility.

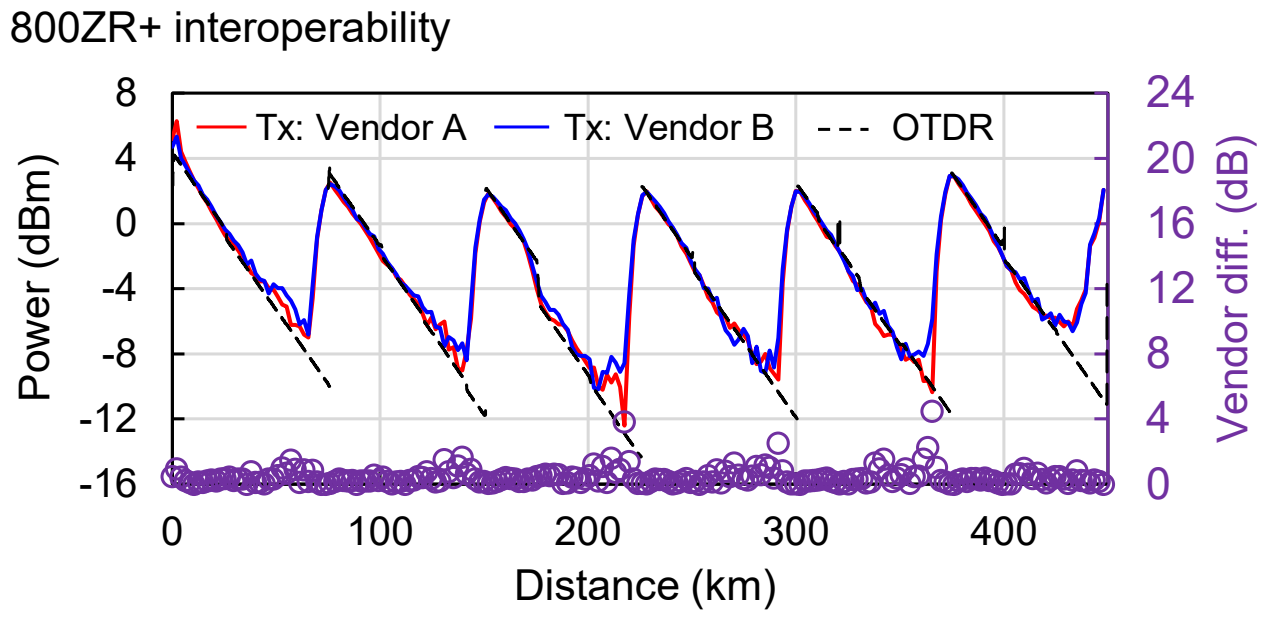

Fig. 19 Multi-vendor interoperability: on-chip LPM results in 800ZR+ (16QAM) mode with transmitter modules from different vendors. Circles represent the difference between the power profiles obtained with the two vendors' transmitters (right vertical axis).

## Acknowledgment

These results were obtained from the commissioned research (NoJPJ012368G60201) by National Institute of Information and Communications Technology (NICT), Japan. We thank the members of NTT Innovative Devices Corporation and VIAVI solutions for their support.